\RequirePackage{lineno} 
\documentclass[reprint,superscriptaddress,twocolumn,showpacs,preprintnumbers,amsmath,amssymb]{revtex4}

\usepackage{graphicx}
\usepackage{dcolumn}
\usepackage{bm}
\usepackage{color}

\def \nobreakseq {\nobreak \hskip 0pt \hbox}

\definecolor{gris25}{gray}{0.75}
\definecolor{violet}{rgb}{0.5,0,0.5}
\definecolor{marron}{rgb}{0.88,0.41,0.}

\begin{document}


\title{Search for Neutrinoless Double-Beta Decay of
  $^{100}$Mo with the NEMO-3 Detector}


\setcounter{footnote}{2}

\author{R.~Arnold}
\affiliation{IPHC, ULP, CNRS/IN2P3\nobreakseq{,} F-67037 Strasbourg, France}
\author{C.~Augier} 
\affiliation{LAL, Univ Paris-Sud\nobreakseq{,} CNRS/IN2P3\nobreakseq{,}
F-91405 Orsay\nobreakseq{,} France}
\author{J.D.~Baker$^{\footnote[2]{Deceased}}$}
\affiliation{Idaho National Laboratory\nobreakseq{,} Idaho Falls, ID 83415, U.S.A.}
\author{A.S.~Barabash}
\affiliation{ITEP, 117218 Moscow, Russia}
\author{A.~Basharina-Freshville} 
\affiliation{UCL, London WC1E 6BT\nobreakseq{,} United Kingdom}
\author{S.~Blondel} 
\affiliation{LAL, Univ Paris-Sud\nobreakseq{,} CNRS/IN2P3\nobreakseq{,}
F-91405 Orsay\nobreakseq{,} France}
\author{S.~Blot}
\affiliation{University of Manchester\nobreakseq{,} Manchester M13
  9PL\nobreakseq{,}~United Kingdom}
\author{M.~Bongrand} 
\affiliation{LAL, Univ Paris-Sud\nobreakseq{,} CNRS/IN2P3\nobreakseq{,}
F-91405 Orsay\nobreakseq{,} France}
\author{V.~Brudanin} 
\affiliation{JINR, 141980 Dubna, Russia}
\author{J.~Busto} 
\affiliation{CPPM, Universit\'e de Marseille\nobreakseq{,} CNRS/IN2P3\nobreakseq{,} F-13288 Marseille\nobreakseq{,} France}
\author{A.J.~Caffrey}
\affiliation{Idaho National Laboratory\nobreakseq{,} Idaho Falls, ID 83415, U.S.A.}
\author{C.~Cerna} 
\affiliation{CENBG\nobreakseq{,} Universit\'e de Bordeaux\nobreakseq{,} CNRS/IN2P3\nobreakseq{,} F-33175 Gradignan\nobreakseq{,} France}
\author{A.~Chapon} 
\affiliation{LPC Caen\nobreakseq{,} ENSICAEN\nobreakseq{,} Universit\'e de
Caen\nobreakseq{,} CNRS/IN2P3\nobreakseq{,} F-14050 Caen\nobreakseq{,} France}
\author{E.~Chauveau} 
\affiliation{University of Manchester\nobreakseq{,} Manchester M13
  9PL\nobreakseq{,}~United Kingdom}
\author{D.~Duchesneau} 
\affiliation{LAPP, Universit\'e de Savoie\nobreakseq{,}
CNRS/IN2P3\nobreakseq{,} F-74941 Annecy-le-Vieux\nobreakseq{,} France}
\author{D.~Durand} 
\affiliation{LPC Caen\nobreakseq{,} ENSICAEN\nobreakseq{,} Universit\'e de
Caen\nobreakseq{,} CNRS/IN2P3\nobreakseq{,} F-14050 Caen\nobreakseq{,} France}
\author{V.~Egorov}
\affiliation{JINR, 141980 Dubna, Russia}
\author{G.~Eurin} 
\affiliation{LAL, Univ Paris-Sud\nobreakseq{,} CNRS/IN2P3\nobreakseq{,}
F-91405 Orsay\nobreakseq{,} France}
\affiliation{UCL, London WC1E 6BT\nobreakseq{,} United Kingdom}
\author{J.J.~Evans} 
\affiliation{University of Manchester\nobreakseq{,} Manchester M13
  9PL\nobreakseq{,}~United Kingdom}
\author{R.~Flack} 
\affiliation{UCL, London WC1E 6BT\nobreakseq{,} United Kingdom}
\author{X.~Garrido} 
\affiliation{LAL, Univ Paris-Sud\nobreakseq{,} CNRS/IN2P3\nobreakseq{,}
F-91405 Orsay\nobreakseq{,} France}
\author{H.~G\'omez} 
\affiliation{LAL, Univ Paris-Sud\nobreakseq{,} CNRS/IN2P3\nobreakseq{,}
F-91405 Orsay\nobreakseq{,} France}
\author{B.~Guillon} 
\affiliation{LPC Caen\nobreakseq{,} ENSICAEN\nobreakseq{,} Universit\'e de
Caen\nobreakseq{,} CNRS/IN2P3\nobreakseq{,} F-14050 Caen\nobreakseq{,} France}
\author{P.~Guzowski} 
\affiliation{University of Manchester\nobreakseq{,} Manchester M13
  9PL\nobreakseq{,}~United Kingdom}
\author{R.~Hod\'{a}k} 
\affiliation{Institute of Experimental and Applied Physics\nobreakseq{,} Czech
Technical University in Prague\nobreakseq{,} CZ-12800
Prague\nobreakseq{,} Czech Republic}
\author{P.~Hubert} 
\affiliation{CENBG\nobreakseq{,} Universit\'e de Bordeaux\nobreakseq{,} CNRS/IN2P3\nobreakseq{,} F-33175 Gradignan\nobreakseq{,} France}
\author{C.~Hugon}
\affiliation{CENBG\nobreakseq{,} Universit\'e de Bordeaux\nobreakseq{,} CNRS/IN2P3\nobreakseq{,} F-33175 Gradignan\nobreakseq{,} France}
\author{S.~Jullian} 
\affiliation{LAL, Univ Paris-Sud\nobreakseq{,} CNRS/IN2P3\nobreakseq{,}
F-91405 Orsay\nobreakseq{,} France}
\author{A.~Klimenko} 
\affiliation{JINR, 141980 Dubna, Russia}
\author{O.~Kochetov} 
\affiliation{JINR, 141980 Dubna, Russia}
\author{S.I.~Konovalov} 
\affiliation{ITEP, 117218 Moscow, Russia}
\author{V.~Kovalenko}
\affiliation{JINR, 141980 Dubna, Russia}
\author{D.~Lalanne} 
\affiliation{LAL, Univ Paris-Sud\nobreakseq{,} CNRS/IN2P3\nobreakseq{,}
F-91405 Orsay\nobreakseq{,} France}
\author{K.~Lang} 
\affiliation{University of Texas at Austin\nobreakseq{,}
  Austin\nobreakseq{,} TX 78712\nobreakseq{,}~U.S.A.}
\author{Y.~Lemi\`ere} 
\affiliation{LPC Caen\nobreakseq{,} ENSICAEN\nobreakseq{,} Universit\'e de
Caen\nobreakseq{,} CNRS/IN2P3\nobreakseq{,} F-14050 Caen\nobreakseq{,} France}
\author{Z.~Liptak} 
\affiliation{University of Texas at Austin\nobreakseq{,}
  Austin\nobreakseq{,} TX 78712\nobreakseq{,}~U.S.A.}
\author{P.~Loaiza} 
\affiliation{Laboratoire Souterrain de Modane\nobreakseq{,} F-73500
Modane\nobreakseq{,} France}
\author{G.~Lutter} 
\affiliation{CENBG\nobreakseq{,} Universit\'e de Bordeaux\nobreakseq{,} CNRS/IN2P3\nobreakseq{,} F-33175 Gradignan\nobreakseq{,} France}
\author{F.~Mamedov}
\affiliation{Institute of Experimental and Applied Physics\nobreakseq{,} Czech
Technical University in Prague\nobreakseq{,} CZ-12800
Prague\nobreakseq{,} Czech Republic}
\author{C.~Marquet} 
\affiliation{CENBG\nobreakseq{,} Universit\'e de Bordeaux\nobreakseq{,} CNRS/IN2P3\nobreakseq{,} F-33175 Gradignan\nobreakseq{,} France}
\author{F.~Mauger} 
\affiliation{LPC Caen\nobreakseq{,} ENSICAEN\nobreakseq{,} Universit\'e de
Caen\nobreakseq{,} CNRS/IN2P3\nobreakseq{,} F-14050 Caen\nobreakseq{,} France}
\author{B.~Morgan} 
\affiliation{University of Warwick\nobreakseq{,} Coventry CV4
7AL\nobreakseq{,} United Kingdom}
\author{J.~Mott} 
\affiliation{UCL, London WC1E 6BT\nobreakseq{,} United Kingdom}
\author{I.~Nemchenok} 
\affiliation{JINR, 141980 Dubna, Russia}
\author{M.~Nomachi} 
\affiliation{Osaka University\nobreakseq{,} 1-1 Machikaney arna
Toyonaka\nobreakseq{,} Osaka 560-0043\nobreakseq{,} Japan}
\author{F.~Nova} 
\affiliation{University of Texas at Austin\nobreakseq{,}
  Austin\nobreakseq{,} TX 78712\nobreakseq{,}~U.S.A.}
\author{F.~Nowacki} 
\affiliation{IPHC, ULP, CNRS/IN2P3\nobreakseq{,} F-67037 Strasbourg, France}
\author{H.~Ohsumi} 
\affiliation{Saga University\nobreakseq{,} Saga 840-8502\nobreakseq{,}
  Japan}
\author{R.B.~Pahlka}
\affiliation{University of Texas at Austin\nobreakseq{,}
  Austin\nobreakseq{,} TX 78712\nobreakseq{,}~U.S.A.}
\author{F.~Perrot} 
\affiliation{CENBG\nobreakseq{,} Universit\'e de Bordeaux\nobreakseq{,} CNRS/IN2P3\nobreakseq{,} F-33175 Gradignan\nobreakseq{,} France}
\author{F.~Piquemal} 
\affiliation{CENBG\nobreakseq{,} Universit\'e de Bordeaux\nobreakseq{,} CNRS/IN2P3\nobreakseq{,} F-33175 Gradignan\nobreakseq{,} France}
\affiliation{Laboratoire Souterrain de Modane\nobreakseq{,} F-73500
Modane\nobreakseq{,} France}
\author{P.~Povinec}
\affiliation{FMFI,~Comenius~Univ.\nobreakseq{,}~SK-842~48~Bratislava\nobreakseq{,}~Slovakia}
\author{Y.A.~Ramachers} 
\affiliation{University of Warwick\nobreakseq{,} Coventry CV4
7AL\nobreakseq{,} United Kingdom}
\author{A.~Remoto}
\affiliation{LAPP, Universit\'e de Savoie\nobreakseq{,}
CNRS/IN2P3\nobreakseq{,} F-74941 Annecy-le-Vieux\nobreakseq{,} France}
\author{J.L.~Reyss} 
\affiliation{LSCE\nobreakseq{,} CNRS\nobreakseq{,} F-91190
  Gif-sur-Yvette\nobreakseq{,} France}
\author{B.~Richards} 
\affiliation{UCL, London WC1E 6BT\nobreakseq{,} United Kingdom}
\author{C.L.~Riddle} 
\affiliation{Idaho National Laboratory\nobreakseq{,} Idaho Falls, ID 83415, U.S.A.}
\author{E.~Rukhadze} 
\affiliation{Institute of Experimental and Applied Physics\nobreakseq{,} Czech
Technical University in Prague\nobreakseq{,} CZ-12800
Prague\nobreakseq{,} Czech Republic}
\author{R.~Saakyan} 
\affiliation{UCL, London WC1E 6BT\nobreakseq{,} United Kingdom}
\author{X.~Sarazin} 
\affiliation{LAL, Univ Paris-Sud\nobreakseq{,} CNRS/IN2P3\nobreakseq{,}
F-91405 Orsay\nobreakseq{,} France}
\author{Yu.~Shitov} 
\affiliation{JINR, 141980 Dubna, Russia}
\affiliation{Imperial College London\nobreakseq{,} London SW7
2AZ\nobreakseq{,} United Kingdom}
\author{L.~Simard} 
\affiliation{LAL, Univ Paris-Sud\nobreakseq{,} CNRS/IN2P3\nobreakseq{,}
F-91405 Orsay\nobreakseq{,} France}
\affiliation{Institut Universitaire de France\nobreakseq{,} F-75005 Paris\nobreakseq{,} France}
\author{F.~\v{S}imkovic} 
\affiliation{FMFI,~Comenius~Univ.\nobreakseq{,}~SK-842~48~Bratislava\nobreakseq{,}~Slovakia}
\author{A.~Smetana}
\affiliation{Institute of Experimental and Applied Physics\nobreakseq{,} Czech
Technical University in Prague\nobreakseq{,} CZ-12800
Prague\nobreakseq{,} Czech Republic}
\author{K.~Smolek} 
\affiliation{Institute of Experimental and Applied Physics\nobreakseq{,} Czech
Technical University in Prague\nobreakseq{,} CZ-12800
Prague\nobreakseq{,} Czech Republic}
\author{A.~Smolnikov} 
\affiliation{JINR, 141980 Dubna, Russia}
\author{S.~S\"oldner-Rembold}
\affiliation{University of Manchester\nobreakseq{,} Manchester M13
  9PL\nobreakseq{,}~United Kingdom}
\author{B.~Soul\'e}
\affiliation{CENBG\nobreakseq{,} Universit\'e de Bordeaux\nobreakseq{,} CNRS/IN2P3\nobreakseq{,} F-33175 Gradignan\nobreakseq{,} France}
\author{I.~\v{S}tekl} 
\affiliation{Institute of Experimental and Applied Physics\nobreakseq{,} Czech
Technical University in Prague\nobreakseq{,} CZ-12800
Prague\nobreakseq{,} Czech Republic}
\author{J.~Suhonen} 
\affiliation{Jyv\"askyl\"a University\nobreakseq{,} FIN-40351 Jyv\"askyl\"a\nobreakseq{,} Finland}
\author{C.S.~Sutton} 
\affiliation{MHC\nobreakseq{,} South Hadley\nobreakseq{,} Massachusetts 01075\nobreakseq{,} U.S.A.}
\author{G.~Szklarz}
\affiliation{LAL, Univ Paris-Sud\nobreakseq{,} CNRS/IN2P3\nobreakseq{,}
F-91405 Orsay\nobreakseq{,} France}
\author{J.~Thomas} 
\affiliation{UCL, London WC1E 6BT\nobreakseq{,} United Kingdom}
\author{V.~Timkin} 
\affiliation{JINR, 141980 Dubna, Russia}
\author{S.~Torre} 
\affiliation{UCL, London WC1E 6BT\nobreakseq{,} United Kingdom}
\author{Vl.I.~Tretyak} 
\affiliation{Institute for Nuclear Research\nobreakseq{,} MSP 03680\nobreakseq{,} Kyiv\nobreakseq{,} Ukraine}
\author{V.I.~Tretyak}
\affiliation{JINR, 141980 Dubna, Russia}
\author{V.I.~Umatov} 
\affiliation{ITEP, 117218 Moscow, Russia}
\author{I.~Vanushin} 
\affiliation{ITEP, 117218 Moscow, Russia}
\author{C.~Vilela} 
\affiliation{UCL, London WC1E 6BT\nobreakseq{,} United Kingdom}
\author{V.~Vorobel} 
\affiliation{Charles University in Prague\nobreakseq{,} Faculty of Mathematics
and Physics\nobreakseq{,} CZ-12116 Prague\nobreakseq{,} Czech Republic}
\author{D.~Waters} 
\affiliation{UCL, London WC1E 6BT\nobreakseq{,} United Kingdom}
\author{A.~\v{Z}ukauskas}
\affiliation{Charles University in Prague\nobreakseq{,} Faculty of Mathematics
and Physics\nobreakseq{,} CZ-12116 Prague\nobreakseq{,} Czech Republic}
\collaboration{NEMO-3 Collaboration}
\noaffiliation


\date{\today}

\begin{abstract}

We report the results of a search for the neutrinoless double-$\beta$ decay
(0$\nu\beta\beta$) of $^{100}$Mo, using the NEMO-3 detector to
reconstruct the full topology of the final state events. With  an exposure of
34.7 kg$\cdot$y, no
evidence for the 0$\nu$$\beta\beta$ signal has been found, yielding a
limit for the light Majorana neutrino mass mechanism of
$T_{1/2}(0\nu\beta\beta)>1.1 \times 10^{24}$ years (90\% C.L.)
once both statistical and systematic uncertainties are taken into
account. Depending on the Nuclear Matrix Elements this corresponds to an upper
limit on the Majorana effective neutrino mass of  $\langle m_{\nu} \rangle <
0.3-0.9$~eV~(90\%~C.L.). Constraints on other lepton number violating mechanisms of
0$\nu\beta\beta$ decays are also given. Searching for high-energy
double electron events in all suitable sources of the detector, no event in the
energy region [3.2-10] MeV is observed for an exposure of 47 kg$\cdot$y.

\end{abstract}


\maketitle


\pagebreak

Many extensions of the Standard Model
  provide a natural framework for neutrino
masses and lepton number violation. 
In particular the see-saw mechanism \cite{Mohapatra1980}, 
which requires the existence of a Majorana neutrino, naturally explains the smallness of neutrino masses. 
The existence of  Majorana neutrinos would also provide a natural
framework for the leptogenesis process~\cite{Leptogenesis} which
could explain the observed baryon-antibaryon asymmetry in the
Universe. The observation of neutrinoless double-$\beta$ decay (0$\nu\beta\beta$)
would prove that neutrinos are Majorana particles \cite{valle} and
that lepton number is not conserved. The isotopes for which a
single-$\beta$ is energetically forbidden or
strongly suppressed are most suitable for the search of this rare
radioactive process. The experimental signature
of 0$\nu \beta \beta$ decays is the emission of two electrons with
total energy ($\mathrm{E_{TOT}}$) equal to the
$\mathrm{Q}$-value of the decay~($\mathrm{Q_{\beta \beta}}$).

The NEMO-3 experiment searches for the double-$\beta$ decay of seven
isotopes by reconstructing the full topology of the final state
events. The NEMO-3 detector \cite{Augier2005}, installed in the Modane underground
laboratory (LSM, France) under a rock overburden of 4800 m.w.e., ran
between February 2003 and January 2011. Here we report on the results
obtained with $^{100}$Mo, the largest sample in NEMO-3, with a mass of 6914
g and $Q_{\beta\beta}= 3034.40 \pm 0.17\ \mathrm{keV}$ \cite{Rahaman2007}. The most stringent previously
published bound on the half-life of the $0\nu\beta\beta$ decay of $^{100}$Mo was
obtained by NEMO-3 using a subset of the data sample analysed here,
placing a limit $\mathrm{T > 4.6 \times 10^{23}}$~y at 90$\%$~C.L.\cite{article0nuphase1}.




The distinctive feature of the NEMO-3 detection method is 
the full reconstruction with 3D-tracking and calorimetric information of
the topology of the final state, comprising two
electrons simultaneously emitted from a common vertex in a $\beta\beta$ source. 
The NEMO-3 detector consists of 20 sectors arranged in a
cylindrical geometry
containing thin (40-60 mg/cm$^2$) source foils of $\beta\beta$
emitters. 
The $^{100}$Mo source foils were constructed from 
either a metallic foil or powder bound by an organic glue to mylar strips
(composite foils). $^{100}$Mo was purified through
physical and chemical processes \cite{Augier2005}. 
The foils are suspended between two concentric cylindrical tracking volumes
consisting of 6180 drift cells operating in Geiger mode
\cite{Augier2005}. To minimize multiple scattering, the gaseous tracking
detector is filled mainly with helium (95\%) with admixtures of ethyl
alcohol (4\%), argon (1\%) and water vapour (0.1\%). The tracking
detector is surrounded by a calorimeter made of large blocks of plastic scintillator (1940 blocks in total)
coupled to low radioactivity 3'' and 5'' diameter photomultiplier tubes
(PMTs). 
The tracking detector, immersed in a magnetic field, is used to identify electron
  tracks and can measure the delay time of any tracks up to
  700 $\mu$s  after the initial event. This is used to tag electron-alpha
($e^- \alpha$) events from the $^{214}$Bi -$^{214}$Po cascade.
The calorimeter measures the energy and the arrival
time of the particles. For $\mathrm{1\ MeV}$ electrons the timing resolution
is $\sigma$ = 250 ps  while the energy resolution is
$\mathrm{FWHM = [14-17] \%/\sqrt{E(MeV)}}$. 
The detector response to the summed energy of the two electrons from
the $0 \nu\beta\beta $ signal is a peak broadened by the energy
resolution of the calorimeter and fluctuations in electron
energy losses in the source foils, which gives a non-Gaussian tail
extending to low energies. The
FWHM of the expected $0 \nu \beta\beta$ two-electron energy spectrum
for $^{100}$Mo is 350~keV. Electrons and photons can
  be identified through tracking
and calorimetry. 
A solenoid surrounding the detector produces a 25 G magnetic field to
reject pair production and external electron events.
The detector is shielded from external gamma rays
by 19 cm of low activity iron and 30 cm of water with boric acid to suppress 
the neutron flux. 
A radon trapping facility was installed at the LSM in autumn 2004,
reducing the radon activity of the air surrounding the
  detector. As a consequence, the radon inside the tracking chamber is
  reduced by a factor of about 6.
The data taken by NEMO-3 are subdivided into two data sets
hereafter referred to as Phase 1 (February 2003 $-$ November 2004) and 
Phase 2 (December 2004 $-$ December 2010), respectively. Results
obtained with both data sets are presented here. 

Twenty calibration tubes located
in each sector near the source foils are used to introduce up to 60 radioactive sources
($^{207}$Bi, $^{232}$U).
The calorimeter absolute energy scale is calibrated every 3 weeks with
$^{207}$Bi sources which provide internal conversion electrons of
482 and 976 keV. The linearity of the PMTs was verified
in a dedicated light injection test during the construction phase and
deviation was found to
be less than 1$\%$ in the energy range [0-4] MeV.
The 1682 keV internal conversion electron peak  of
$^{207}$Bi is
used to determine the systematic uncertainty on
  the energy scale:
the data-Monte Carlo disagreement in reconstructing the peak position is less than
0.2$\%$. For 99$\%$ of the PMTs the energy
scale is known with an accuracy better than
2$\%$. Only these PMTs are used in the data analysis 
presented here. The relative gain and time variation of individual PMTs is surveyed twice a day by a
light injection system; PMTs that show a gain variation of more than
5\% compared to a linear interpolation between two successive absolute
  calibrations with $^{207}$Bi are rejected from the analysis.



\begin{figure}[htb]
\includegraphics[scale=0.5]{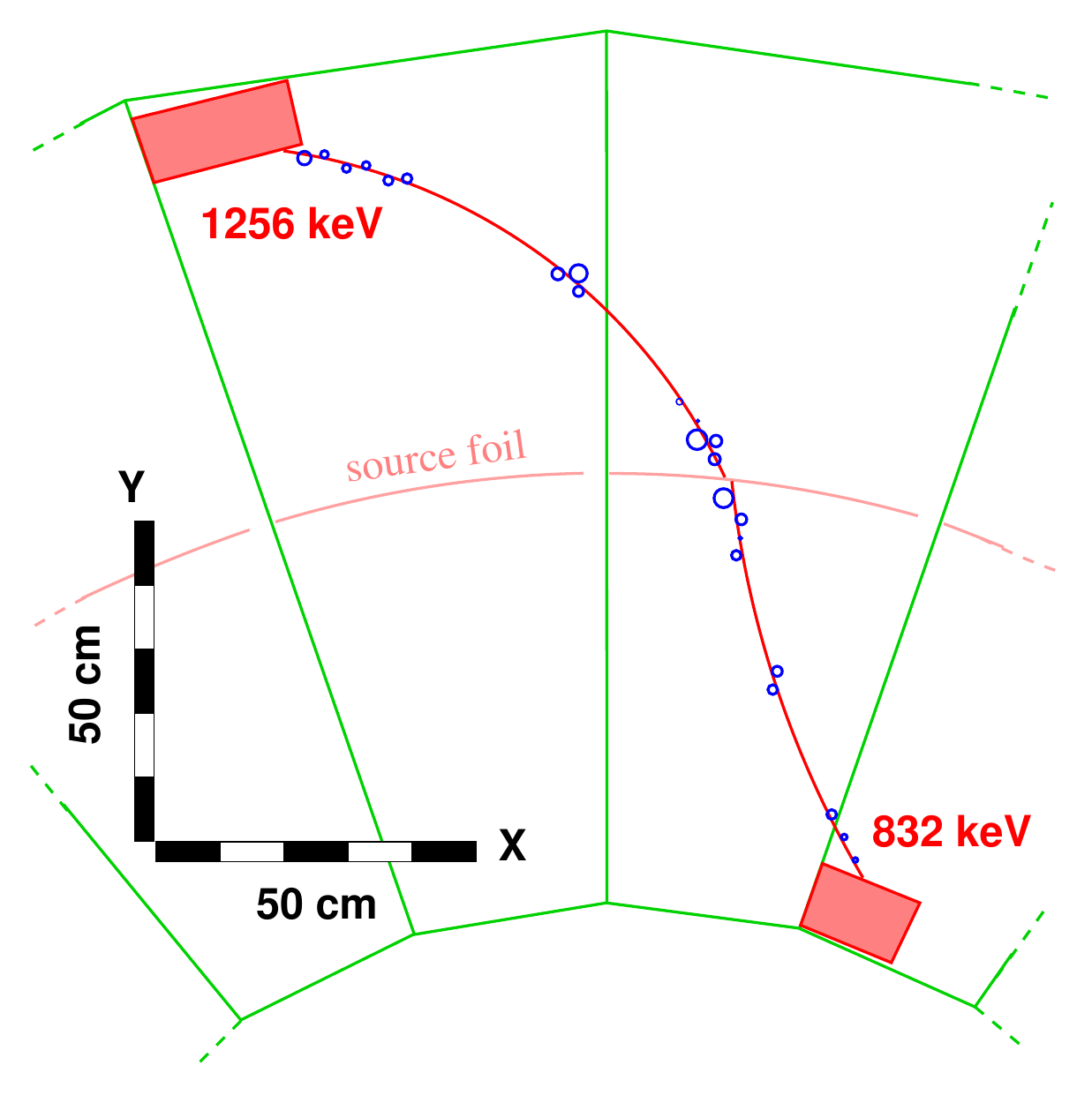}
\caption{\label{fig:event} Transverse view of a reconstructed $\beta\beta$ data event with a two electron energy sum of 2088 keV.  Two
    tracks are reconstructed from a single vertex in the
    source foil, with an electron-like curvature in the magnetic field, and are each
    associated to an energy deposit  in a calorimeter block. 
}
\end{figure}

Two-electron (2e$^{-}$) events are selected with the following
requirements. Two tracks with a length greater than 50 cm
and an electron-like curvature
must be reconstructed. The geometrical efficiency is
  28.3 $\%$.
Both tracks
are required to
originate from a common reconstructed vertex in the $^{100}$Mo source
foil with transverse and longitudinal resolutions of
  $\mathrm{\sigma_t=[2-3]}$ mm and $\mathrm{\sigma_l=[7-13]}$ mm,
  respectively \cite{Augier2005}. The tracks terminate in isolated scintillator blocks with a single
energy deposit greater than 0.2 MeV. A time of
  flight criterion requires that the two electrons should be emitted
  from the source foil. There must be no
  photons or delayed tracks present in the event. Fig.~\ref{fig:event} shows a candidate $\beta\beta$ event observed in the data.



When searching for rare processes, the background estimation is of
paramount importance.  An exhaustive program, described in detail in \cite{backgroundpaper}, has been carried out to
measure the backgrounds in the NEMO-3 detector.
The sources of backgrounds relevant to the $0 \nu \beta \beta$ search in $^{100}$Mo
are the irreducible background from its $2 \nu \beta \beta$ decay, as
well as the decays of  $^{214}$Bi and $^{208}$Tl originating from the natural decay
chains of $^{238}$U and $^{232}$Th. 
These high-$\mathrm{Q_\beta}$ isotopes can produce
2e$^-$ events by one of the following mechanisms:
a $\beta$ decay followed by a M{\o}ller
scattering interaction, a $\beta$-$\gamma$ cascade followed by a Compton
scattering of the emitted $\gamma$ close to the vertex or 
a $\beta$ decay accompanied by emission of an internal conversion electron.
The background isotopes can either be
present in the $\beta \beta$ source foils, or can result from $^{222}$Rn or $^{220}$Rn
emanation. The isotopes $^{214}$Bi and $^{208}$Tl are progenies of $^{222}$Rn and
$^{220}$Rn respectively and can end up on the surfaces of the source
foils and drift cell wires located in the vicinity of the foils. 

The $^{100}$Mo foil internal backgrounds are measured with the
full statistics of the entire data set using topologies, energy and
timing information specific for the background in question
\cite{backgroundpaper}. The results are summarized in Table
\ref{tab:results_background}. The $^{214}$Bi activity inside the source foils and from $^{222}$Rn
is measured with the $\mathrm{1 e^{-}\alpha}$ delayed coincidence channel
which is an efficient and background free signature of the $^{214}$Bi-$^{214}$Po
$\beta$-$\alpha$ decay cascade. The total $^{222}$Rn activity inside the 28 m$^{3}$ tracker chamber
is measured to be  1138 $\pm$ 199  mBq and 205 $\pm$ 77 mBq in Phase
1 and Phase 2, respectively. The $^{220}$Rn activity is found to be at
a level of 3 mBq giving a negligible contribution to  2e$^-$ events.
The non-uniform distribution of the deposition
  of the Rn daughters
inside the tracker is also taken into account \cite{backgroundpaper}.  The $^{214}$Bi
location inside the source foils, or on the surface of the foils and
drift wires can be statistically separated by fitting the $\alpha$ track length
distribution. 
The $^{208}$Tl activity is measured with the
$\mathrm{1 e^-\ n\gamma}$ channel
(n$\geq$1), which contains events due to the $\beta$
decay of $^{208}$Tl followed
by de-excitation $\gamma$-rays of the $^{208}$Pb daughter isotope. The $^{214}$Bi and $^{208}$Tl contamination measurements
are independently verified using the 2e$^{-}$ + X event topologies
 in the energy range [2.8-3.2] MeV where
a large part of the $^{100}$Mo 0$\nu\beta\beta$ signal is expected. 
For $^{214}$Bi, 6 events with a 2e$^-$1$\alpha$ topology are observed in the data after a total 
exposure of 34.7 kg$\cdot$y while 9.4 $\pm$ 0.5 are expected from simulations.
For $^{208}$Tl, 7 events with a 2e$^-$n$\gamma$ topology are
observed in the same data sample, while 8.8 $\pm$ 0.6
  are
expected. Both tests, although statistically limited, show that the
prediction of the background contribution to 0$\nu \beta\beta$ from $^{214}$Bi
and $^{208}$Tl are reliable within the quoted uncertainties. 

Neutrons produced by ($\alpha$,n) reactions and spontaneous fission
reactions are also a potential source of background. They can be
thermalized in the scintillator material and subsequently captured
producing $\gamma$-rays with energies up to $\mathrm{10\ MeV}$. These high energy
$\gamma$-rays can interact with the source foils and mimic $\beta\beta$
events through pair creation, double Compton scattering or a Compton
scattering followed by a M{\o}ller interaction. A model of the neutron background
is validated with dedicated runs using a calibrated
Am-Be neutron source and is found to be negligible for the
0$\nu\beta\beta$ search.

\begin{table}
\caption{\label{tab:results_background} Foil contamination
  activities for the $^{100}$Mo source measured by NEMO-3, and their contribution to the
  background in the 2e$^{-}$ channel within the $\mathrm{E_{TOT}}$ = [2.8-3.2] MeV range. The uncertainties are statistical only. The
  masses of $^{100}$Mo from metallic and composite sources are
  respectively 2479 g and 4435 g.}
\begin{ruledtabular}
\begin{tabular}{c|cc}
 & \multicolumn{2}{c}{$^{100}$Mo metallic/composite} \\
& Activities ($\mu$Bq/kg) & Number of 2e$^{-}$ events \\
 \hline
      \rule{0pt}{2.4ex} $^{214}$Bi internal         &   60 $\pm$ 20/380 $\pm$ 40 & 0.07 $\pm$ 0.02/0.91 $\pm$ 0.07   
      \\
      $^{208}$Tl internal               & 87 $\pm$ 4/128      $\pm$ 3 & 0.91 $\pm$ 0.04/2.39 $\pm$ 0.06
      \\
\end{tabular}
\end{ruledtabular}
\end{table}

The radon and external background model are verified in the energy region
close to the $^{100}$Mo Q$_{\beta\beta}$-value by selecting 2e$^{-}$
events from the sectors containing Cu, $^{130}$Te and natural
Te foils. The internal contamination of these foils with radioactive
isotopes is independently measured. The 2$\nu\beta\beta$
decay of $^{130}$Te gives no contribution above 2.4
MeV \cite{articleTe130}. With an exposure of $\mathrm{13.5\ kg\cdot y}$ only 3 
events with 2e$^{-}$ from the sectors containing Cu, $^{130}$Te and natural
Te foils remain in the energy window
[2.8-3.2] MeV in the full data set, compared to a MC expectation of 3.6 $\pm$
0.2 events, dominated by radon background.  

\begin{figure}
\includegraphics[scale=0.46]{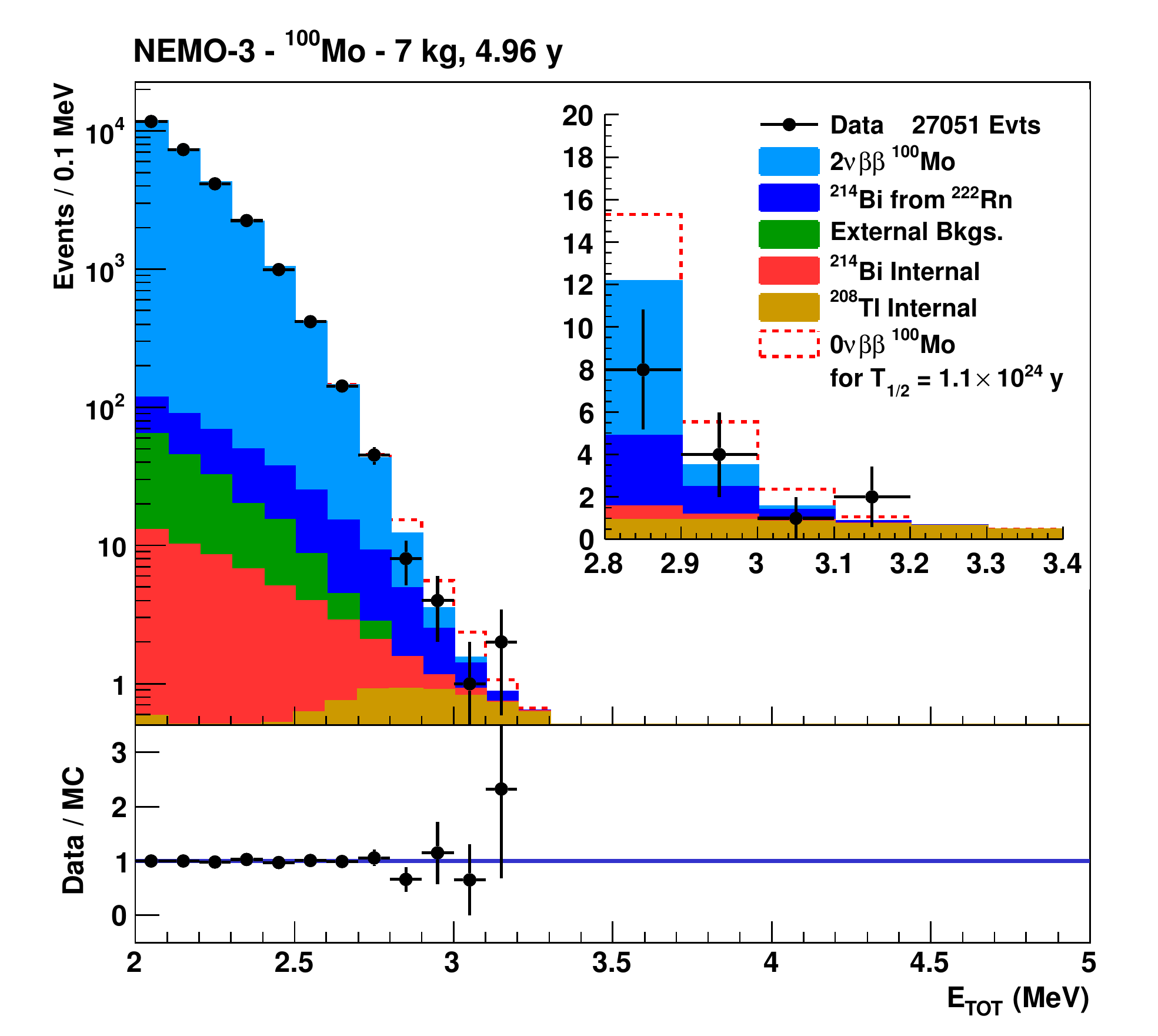}
\caption{\label{fig:mobb2nu}  Distribution of the two-electron energy
  sum, E$_{\mathrm{TOT}}$ and the ratio between the observed and the
  expected distributions from Monte Carlo simulations.
The  distribution is for E$_{\mathrm{TOT}} > $ 2 MeV and is
  obtained with an exposure of 34.7 kg$\cdot$y. 
The solid histogram represents the expected spectrum consisting of
2$\nu\beta\beta$ decays and radioactive backgrounds determined by
Monte~Carlo simulations.
The dashed histogram in the E$_{\mathrm{TOT}}$ distribution
represents a hypothetical
0$\nu\beta\beta$ signal corresponding to a half-life of
  $1.1 \times 10^{24}$ y.
}
\end{figure}

Another background to 0$\nu\beta\beta$ decays is the 2$\nu\beta\beta$
decay allowed in the Standard Model. There are 683\nobreakseq{,}049
2e$^{-}$ events in the full energy range of E$_{\mathrm{TOT}}$ =
[0.4 - 3.2] MeV with a signal-to-background ratio of 76. The 2$\nu\beta\beta$ contribution is found by
fitting the energy sum distribution  of two electrons using the shape of the 2$\nu\beta\beta$
spectrum predicted by the Single State Dominance model for $^{100}$Mo
\cite{ssd}, taking into account
the backgrounds quoted
  previously. Figure~\ref{fig:mobb2nu} shows for $\mathrm{E_{TOT} > 2\ MeV}$ the spectra of the 2e$^{-}$
energy sum, exhibiting good agreement between
the data and MC. The number of 2$\nu\beta\beta$ events obtained from this fit
with E$\mathrm{_{TOT}}$ $>$ 2 MeV corresponds to a $^{100}$Mo half-life of 
$\mathrm{T_{1/2}(2\nu\beta\beta) = [ 6.93 \pm
  0.04\ (stat) ] \times 10^{18}}$~y, in agreement with the previously published result
\cite{article0nuphase1} and with \cite{PDG}. 

Figure ~\ref{fig:mobb2nu} shows the tail of the $\mathrm{E_{TOT}}$ distribution in the 
energy window $\mathrm{E_{TOT}}$ = [2.8-3.2] MeV around the Q$_{\beta\beta}$-value of
$^{100}$Mo 0$\nu \beta\beta $ decay.
The
background contributions in this energy window are shown in Table
\ref{tab:background}. No events are observed in the region of $\mathrm{E_{TOT}}$ = [3.2-10] MeV for 
  NEMO-3 sources containing isotopes with Q$_{\beta\beta}$-value below 3.2 MeV
  ($^{100}$Mo, $^{82}$Se, $^{130}$Te, $^{116}$Cd) or without $\beta
\beta$ emitter isotopes (Cu) during the
  entire running period, which corresponds
to an exposure of $\mathrm{47\ kg\cdot y}$.

\begin{table}
\caption{\label{tab:background} Number of expected background and
  observed 2e$^-$ events in Phase 1 and Phase 2
  after a 34.7 kg$\cdot$y exposure with
  $^{100}$Mo in the $\mathrm{E_{TOT}}$ = [2.8-3.2] MeV range. 
The $0 \nu \beta
\beta$ detection efficiency is 4.7$\%$ in this window.}
\begin{ruledtabular}
\begin{tabular}{ccccc}
Data sets & Phase 1 & Phase 2 & Combined \\
\hline
External background                & $<$ 0.04 & $<$ 0.16 &      $<$ 0.2         \\
$^{214}$Bi from $^{222}$Rn                   & 2.8 $\pm$ 0.3 & 2.5 $\pm$ 0.2 & 5.2 $\pm$ 0.5      \\
$^{214}$Bi internal      & 0.20 $\pm$ 0.02  & 0.80 $\pm$ 0.08 & 1.0 $\pm$ 0.1   \\
$^{208}$Tl internal                  & 0.65 $\pm$ 0.05 & 2.7 $\pm$ 0.2 & 3.3 $\pm$ 0.3  \\
$2 \nu \beta \beta$                  & 1.28 $\pm$ 0.02 & 7.16 $\pm$
0.05 & 8.45 $\pm$ 0.05  \\
Total expected           & 4.9 $\pm$ 0.3  & 13.1 $\pm$ 0.3 & 18.0 $\pm$ 0.6    \\
\hline
Data & 3 & 12 & 15 \\
\end{tabular}
\end{ruledtabular}
\end{table}


As no event excess is observed in the data above the background
expectation, a limit on the 0$\nu\beta\beta$ decay of $^{100}$Mo is set. 
The systematic uncertainties that are used in setting the
0$\nu\beta\beta$ limit have two main components, the uncertainty on
the 0$\nu\beta\beta$ detection efficiency and the uncertainties on the
background contribution. The uncertainty on the signal efficiency is
determined using dedicated runs with activity-calibrated $^{207}$Bi sources and
is found to be 7$\%$.
The systematic uncertainties on the background contributions are
due to  the activities of 2$\nu\beta\beta$, $^{214}$Bi and $^{208}$Tl. 
The uncertainty on 2$\nu\beta\beta$ is obtained from the fit to
2e$^{-}$ events above 2 MeV described above and is 0.7\%. 
The uncertainty on the $^{214}$Bi
contribution from $^{222}$Rn and internal foil contamination,
estimated by comparing the activities of this isotope
measured independently in 1e1$\alpha$ and 1e1$\gamma$ channels, is
10\%. The uncertainty on the $^{208}$Tl contamination is
determined from dedicated runs with a calibrated $^{232}$U source and
is found to be 10\%. As a result of these estimates, a
systematic uncertainty of 10\%  on the background
contribution from $^{214}$Bi and
$^{208}$Tl radioactive impurities is assumed in setting the limit on
the $^{100}$Mo 0$\nu\beta\beta$  decay.

\begin{table}
\caption{\label{tab:limits} Limits at 90$\%$C.L. on the half-lives of
  lepton number violating  processes (in units of $10^{24}$ y).}
\begin{ruledtabular}
\begin{tabular}{ccccc}
0$\nu\beta\beta$ Process & Statistical  & With & Expected with
 
\\
 &  only & systematics & systematics (median,
\\
 &  & & [-1 $\sigma$,+1 $\sigma$] range)
\\
\hline
Mass mechanism & 1.1 & 1.1 & 1.0 [0.7-1.4] \\
RH current $\langle \lambda \rangle$ & 0.7 & 0.6 & 0.5 [0.4-0.8] \\
RH current $\langle \eta \rangle$ & 1.0 & 1.0 & 0.9 [0.6-1.3] \\
Majoron & 0.050 & 0.044 & 0.039 [0.027-0.059] \\
\end{tabular}
\end{ruledtabular}
\end{table}

The limit is set using a modified frequentist analysis that employs a
log-likelihood ratio test statistic \cite{CLs}. The
method uses the full information of the binned energy sum distribution
in the $\mathrm{E_{TOT}}$ = [2.0-3.2] MeV energy range
for signal and background (Figure~\ref{fig:mobb2nu}), as well as
the statistical and systematic uncertainties and their correlations
as described in more detail in \cite{CLs,Nd-PRC}. The data are
described well by the background-only hypothesis with 
1-CL$_{b}$ = 64.7 \%, where 1-CL$_{b}$ is the p-value of the
background-only hypothesis. 

The $0 \nu \beta
\beta$ detection efficiency for $^{100}$Mo in NEMO-3 is 11.3$\%$ for the
energy sum of two electrons above 2 MeV. Taking into account the total exposure of
34.7 kg$\cdot$y, a limit on the light Majorana neutrino mass
mechanism for 0$\nu\beta\beta$ decay of $^{100}$Mo is set (Table
\ref{tab:limits}). 
The result agrees with the expected sensitivity of
the experiment and is twice more stringent
than the previous best limit for this isotope \cite{article0nuphase1}. 
The corresponding upper limit on the effective Majorana neutrino mass
is $\langle m_{\nu} \rangle < 0.3-0.9$~eV (90\% C.L.), where the range is
determined by existing uncertainties in
the nuclear matrix 
element (NME) 
\cite{Suhonen2012,Simkovic2013,Iachello2009,Rath2010,Rodriguez2010}
and phase space \cite{Iachello2012,Stoica2013} calculations.

\begin{table*}[htb]
\caption{\label{tab:comparison} Limits at 90$\%$ C.L. on the half-lives and lepton number
  violating parameters. $\langle m_{\nu} \rangle$   contraints, both
  experimental published values and recalculated from \cite{Suhonen2012,Simkovic2013,Iachello2009,Rath2010,Rodriguez2010,Menendez2009}  are shown.}
\begin{ruledtabular}
\begin{tabular}{ccccccccc}
Isotope & Half-life & $\langle m_{\nu} \rangle$   & $\langle m_{\nu} \rangle$  & $\langle \lambda
\rangle$   &
$\langle \eta \rangle$   & $\mathrm{\lambda^{'}_{111}/f}$
 & $\langle g_{ee} \rangle$  \\ 
& ($\mathrm{10^{25}}$ y) & (eV)   &  (eV) & ($\mathrm{10^{-6}}$)  & ($\mathrm{10^{-8}}$)   & ($\mathrm{10^{-2}}$)   & ($\mathrm{10^{-5}}$)   \\
& published & published & recalculated & published & published & published & published \\
\hline
$^{100}$Mo This Work & 0.11 & 0.33-0.87 & 0.33-0.87 & 0.9-1.3$\mathrm{^a}$ & 0.5-0.8$\mathrm{^a}$ & 4.4-6.0 & 1.6-4.1$\mathrm{^a}$ \\
$^{130}$Te \cite{cuoricino,cuoricino2} & 0.28 & 0.3-0.71 & 0.31-0.75 & 1.6-2.4$\mathrm{^b}$ & 0.9-5.3$\mathrm{^b}$ & & 17-33$\mathrm{^c}$ \\
$^{136}$Xe \cite{kamlandzen,KZ_majoron} & 1.9 & 0.14-0.34 & 0.14-0.34 & & & & 0.8-1.6 \\
$^{76}$Ge \cite{gerda} & 2.1 & 0.2-0.4 & 0.26-0.62 & & & & \\
$^{76}$Ge \cite{heidelbergmoscow,HeidelbergMoscowRightCurrent} & 1.9 & 0.35  & 0.27-0.65 & 1.1 & 0.64 & & 8.1 \\
\end{tabular}
\end{ruledtabular}
$\mathrm{^a}$ obtained with half-lives given in Table III, $\mathrm{^b}$ using the half-life limit of 2.1$\times 10^{23}$y, $\mathrm{^c}$ using the half-life limit 
of 2.2$\times 10^{21}$y.
\end{table*}

Constraints on other lepton number violating mechanisms of
0$\nu\beta\beta$ are set. Right-left symmetric models can give rise to 0$\nu\beta\beta$ due to
the presence of right-handed currents  in the electroweak
Lagrangian. This mechanism leads to different angular and single energy
distributions of the final state electrons and can therefore be
distinguished from other mechanisms in a NEMO-like experiment
\cite{SNemo}. The corresponding half-life limits are given in Table
\ref{tab:limits} and translate into an upper bound on the coupling between
right-handed quark and lepton currents of
$\langle \lambda \rangle < (0.9-1.3) \times
10^{-6}$ (90\% C.L.);
and into an upper bound on the coupling between
right-handed quark and left-handed lepton currents of
$\langle \eta \rangle < (0.5-0.8) \times 10^{-8}$
(90\% C.L.). The constraints are
obtained using the NME calculations from
\cite{Suhonen2002,Tomoda1991,Muto1989}.

In supersymmetric models the 0$\nu\beta\beta$ process
can be mediated by a gluino or neutralino exchange. Using the above
half-life limit and the NME from \cite{MatrixElementsSupersymmetry} an upper bound
is obtained on the trilinear $R$-parity violating supersymmetric
coupling of $\mathrm{\lambda^{'}_{111} < (4.4-6.0) \times 10^{-2} f}$,
 where $\mathrm{f = \left(\frac{M_{\tilde{q}}}{1~TeV}
\right)^2\left(\frac{M_{\tilde{g}}}{1~TeV} \right)^{1/2}}$ and $\mathrm{M_{\tilde{q}}}$ and $\mathrm{M_{\tilde{g}}}$ represent the squark and the
gluino masses.

Finally, the 0$\nu\beta\beta$  decay can be accompanied by a light or
massless boson that weakly couples to the neutrino, called a Majoron. In
this case the energy sum of the two emitted electrons, $\mathrm{E_{TOT}}$, will
have a broad spectrum ranging from zero to $Q_{\beta\beta}$ of $^{100}$Mo. The
exact shape will depend on the spectral index $\mathrm{n}$, which determines the
phase space dependence on the energy released in the decay, 
$\mathrm{G^{0\nu} \propto (Q_{\beta\beta}-E_{TOT})^{n}}$. The lower
bound on the half-life of the Majoron accompanied 0$\nu\beta\beta$
decay with the spectral index $\mathrm{n=1}$ is given in Table \ref{tab:limits}. This is almost a factor of two more
stringent limit than the previous best limit for this isotope
\cite{majoron}. Taking into account the phase
  space factors given in \cite{Doi1988} and the NME calculated in
\cite{Suhonen2012,Simkovic2013,Iachello2009,Rath2010,Rodriguez2010}, an upper bound on the Majoron-neutrino coupling constant
has been obtained, $\langle g_{ee} \rangle < (1.6-4.1) \times
10^{-5}$. 

All the obtained limits on the lepton number violating parameters are comparable with the best current results
obtained with other isotopes, as shown in Table \ref{tab:comparison}.  

In summary, with an exposure of 34.7 kg$\cdot$y, no evidence for the
0$\nu \beta \beta$ decay of $^{100}$Mo is found. Taking
into account statistical and systematic uncertainties, the half-life
limit for the light Majorana neutrino mass mechanism is
 $T_{1/2}(0\nu\beta\beta)>1.1 \times 10^{24}$ years (90\% C.L.). The
 corresponding limit on the effective Majorana neutrino mass is $\langle
m_{\nu} \rangle < 0.3-0.9$~eV, depending on the nuclear
matrix elements assumed.
The absence of a constant background in the high energy part of the
spectrum is an encouraging result for future 0$\nu\beta\beta$
NEMO-3 like
experiments that plan to use high Q$_{\beta\beta}$-value isotopes such as $^{48}$Ca,
$^{96}$Zr and $^{150}$Nd. 

We thank the staff of the Modane Underground Laboratory for its
technical assistance in running the experiment. We acknowledge
support by the grants agencies of the Czech
Republic, CNRS/IN2P3 in France, RFBR in Russia, STFC
  in U.K. and NSF in U.S.




\end{document}